\documentclass[twocolumn,showpacs,amsmath,amssymb]{revtex4}

\usepackage{graphicx}
\usepackage{dcolumn}
\usepackage{bm}
\begin{document}

\bibliographystyle{osa}
\title{Experimental realization of a ballistic spin interferometer based on the 
Rashba effect using a nanolithographically defined square loop array}
\author{Takaaki Koga$^{1,2,3}$}
\email{koga@ist.hokudai.ac.jp}
\altaffiliation[Present address: ]{Graduate School of Information Science 
and  Technology, Hokkaido University, Sapporo, 060-0814, Japan}
\author{Yoshiaki Sekine$^{2}$}
\author{Junsaku Nitta$^{2,3}$}
\altaffiliation[Present address: ]{Graduate School of Engineering, 
Tohoku University, Sendai, 980-8579, Japan}
\affiliation{$^1$PRESTO, Japan Science and Technology Agency, 4-1-8, Honchou, 
Kawaguchi, Saitama, 332-0012, Japan\\
$^2$NTT Basic Research Laboratories, NTT Corporation,\\ 
3-1, Morinosato-Wakamiya, Atsugi, Kanagawa, 243-0198, Japan\\
$^3$CREST, Japan Science and Technology Agency, 4-1-8, Honchou, Kawaguchi, 
Saitama, 332-0012, Japan}


\begin{abstract}
The gate-controlled electron spin interference was observed in 
nanolithographically defined square loop (SL) arrays fabricated using 
In$_{0.52}$Al$_{0.48}$As/In$_{0.53}$Ga$_{0.47}$As/In$_{0.52}$Al$_{0.48}$As 
quantum wells. In this experiment, we demonstrate electron spin precession in 
quasi-one-dimensional channels that is caused by the Rashba effect. It turned 
out that the spin precession angle $\theta$ was gate-controllable by more 
than 0.75$\pi$ for a sample with $L=1.5\mu$m, where $L$ is the side length of 
the SL. 
Large controllability of $\theta$ by the applied gate voltage as such is a 
necessary requirement for the realization of the spin FET device proposed by 
Datta and Das [Datta {\it et. al.}, Appl. Phys. Lett. {\bf 56}, 665 
(1990)] as well as for the manipulation of spin qubits using the Rashba 
effect.
\end{abstract}

\maketitle

Exploitation of spin degree of freedom for the conduction carriers provides a 
key strategy for finding new functional devices in semiconductor 
spintronics \cite{awschalombook02,datta90,koga02filter,nitta99,bercioux04,kato05}. 
A promising approach for manipulating spins in semiconductor 
nanostructures is the utilization of spin-orbit (SO) 
interactions. In this regard, lifting of the spin degeneracy in the 
conduction (or valence) band due to the structural inversion asymmetry is 
especially called the ``Rashba effect'' \cite{rashba60,bychkov84}, the 
magnitude 
of which can be controlled by the applied gate voltages and/or specific 
design of the sample heterostructures \cite{nitta97,koga02wal}.

Recently, we proposed a ballistic spin interferometer (SI) using a square 
loop (SL) geometry, where an electron spin rotates by an angle 
$\theta$ due to the Rashba effect as it travels along a side of the SL 
ballistically \cite{koga04}. 
In a simple SI model, an incident electron wave to the SI (see Fig. 1 
in Ref.~\onlinecite{koga04}) is split by a ``hypothetical'' beam 
splitter into two partial waves, where each of these partial waves follows 
the SL path in the clockwise (CW) and counter-clockwise (CCW) directions, 
respectively. 
Then, they interfere with each other when they come back to the incident point 
(at the beam splitter). As a consequence, the incident electron would either 
scatter back on the incident path (called ``path1'') or emerge 
on the other path (called ``path2''). The backscattering probability to path1 
($P_{\rm back}$) for the case that the incident electron is spin unpolarized 
is given by \cite{koga04}, 
\begin{eqnarray}
\begin{array}{ll}
P_{\rm back}&=\frac{1}{2}+\frac{1}{4}\left ( {\rm cos}^4\theta + 
4{\rm cos}\theta {\rm sin}^2 \theta + {\rm cos}2\theta \right ) 
{\rm cos}\phi\\
&\equiv \frac{1}{2}+A(\theta){\rm cos}\phi,\label{eq:A_theta}
\end{array}
\end{eqnarray}
where $\phi$ is the quantum mechanical phase due to the vector potential 
responsible for the magnetic field ${\bf B}$ piercing the SL 
($\phi=2eBL^2/\hbar$, $L$ being the side length of the SL) and $\theta$ 
is the spin precession angle when the electron propagates through each side of 
the SL due to the Rashba effect ($\theta=2\alpha m^* L/\hbar^2$, $\alpha$ and 
$m^*$ being the Rashba SO coupling constant and the electron effective mass, 
respectively). A plot of $A(\theta)$ as a function of $\theta$ is found in 
Ref.~\onlinecite{koga04}. We note that $A(\theta)$ corresponds to the 
amplitude of the Al'tshuler-Aronov-Spivak(AAS)-type oscillation of electric 
conductance experimentally \cite{altshuler81}. Equation~(\ref{eq:A_theta}) 
predicts that the amplitude of the AAS oscillation should be modulated as a 
function of $\theta$, which, in turn, can be controlled by the applied gate 
voltage $V_{\rm g}$ through the variation of the $\alpha$ values.

\begin{figure}
\includegraphics[width=7cm,clip,keepaspectratio]{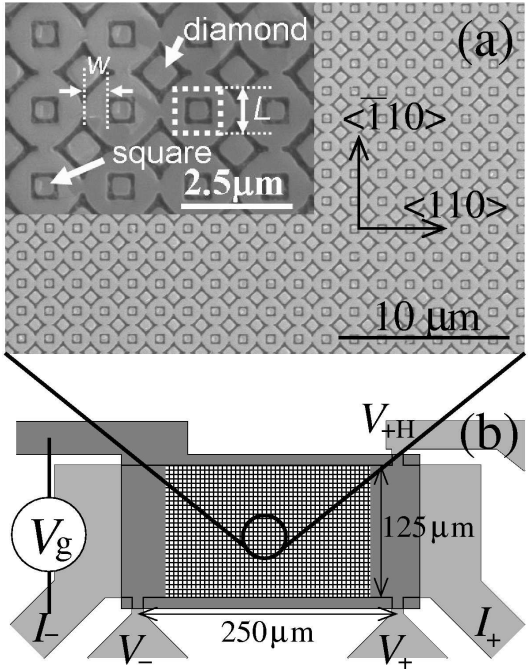}
\caption{\label{fig:interferometer_sample} (a) SEM micrographs of the 
nanolithographycally defined square loop array ($L=1.2\mu$m). A 
two-dimensional electron gas exists in the relatively light regions. 
(b) Schematic diagram for the Hall bar sample used in the present 
experiment.}
\end{figure}

In this Letter, we present the first experimental demonstration of the 
SI using nanolithographically defined SL arrays in epitaxially grown (001) 
In$_{0.52}$Al$_{0.48}$As/In$_{0.53}$Ga$_{0.47}$As/In$_{0.52}$Al$_{0.48}$As 
quantum wells (QW). Details of the sample preparation are following: 
we use the same MOCVD-grown epi-wafers of 
In$_{0.52}$Al$_{0.48}$As/In$_{0.53}$Ga$_{0.47}$As/In$_{0.52}$Al$_{0.48}$As 
QWs as those we used for the weak antilocalization (WAL) study previously 
(samples1-4 in Ref.~\onlinecite{koga02wal}). We first exploit the electron 
beam lithography (EBL) and electron cyclotron resonance (ECR) plasma etching 
techniques to define an array of SLs in the area of 150$\times$200 $\mu$m$^2$. 
We then use the photolithography and wet etching techniques to form a Hall 
bar mesa of the size of 125$\times$250 $\mu$m$^2$ over the SL array 
regions. In this way, the area of the final SL array region in the Hall bar 
mesa is 125$\times$200 $\mu$m$^2$ 
[see Fig.~\ref{fig:interferometer_sample}(b)]. 
These samples have a gate electrode (Au) covering the entire Hall bar, 
using a 100 nm thick SiO$_2$ layer as a gate insulator, which makes it 
possible to control the sheet carrier density $N_{\rm S}$ and the 
Rashba spin-orbit parameter $\alpha$ by the applied gate voltage $V_{\rm g}$. 
We note that all the measurements were carried out at $T =$ 0.3 K using a 
$^3$He cryostat, exploiting the conventional ac lock-in technique. When 
the electric sheet conductivities $\sigma_{\rm 2D}$ of these 
samples were measured [using the electrodes labeled by $I_+$, $I_-$, $V_+$ 
and $V_-$ in Fig.~\ref{fig:interferometer_sample}(b)] as a function of 
$B$ (${\bf B}\perp$ to the sample 
surface) for a given $V_{\rm g}$ [denoted as $\sigma_{\rm 2D}(B)$], the 
Hall voltages were also measured using the electrodes labeled by 
$V_+$ and $V_{+\rm H}$. In this way, we were able to 
monitor $\sigma_{\rm 2D}(B)$ and $N_{\rm S}$ at the same time 
for each given $V_{\rm g}$. We then investigate the amplitude of the AAS 
oscillations at $B=0$ [denoted as $\Delta\sigma_{\rm 2D}(B=0)$], as a 
function of $V_{\rm g}$ (equivalently $N_{\rm S}$), to test the prediction 
of the SI \cite{koga04}. 

Examples of the scanning electron micrographs (SEM) of the SL pattern 
used in the present experiment are shown in 
Fig.~\ref{fig:interferometer_sample}(a). We note that electrons exist 
in the relatively lighter regions of the picture. The relatively darker 
lines and curves that define the ``diamond'' ({\LARGE $\diamond$}) and 
``square'' ($\square$) 
shapes in Fig.~\ref{fig:interferometer_sample}(a), 
are the dry-etched regions by the ECR plasma etching. We note that electrons 
exist in these diamond- and square-shaped islands. However, these islands 
do not contribute to the electric conductivity, since they are not 
electrically connected one another. We sketch a SL path for the spin 
interference by the dotted white square in the inset of 
Fig.~\ref{fig:interferometer_sample}(a), 
where electrons would be localized if the type of the spin interference 
is constructive. The width $W$ of the SL path is also defined 
in Fig.~\ref{fig:interferometer_sample}(a). We used $W=0.5\mu$m 
throughout the present experiment. We can see that these SLs are 
electrically connected with the neighboring SLs. As a result, they 
contribute to the electric conductivity of the whole Hall bar. 
 
\begin{figure}
\includegraphics[width=7cm,clip,keepaspectratio]{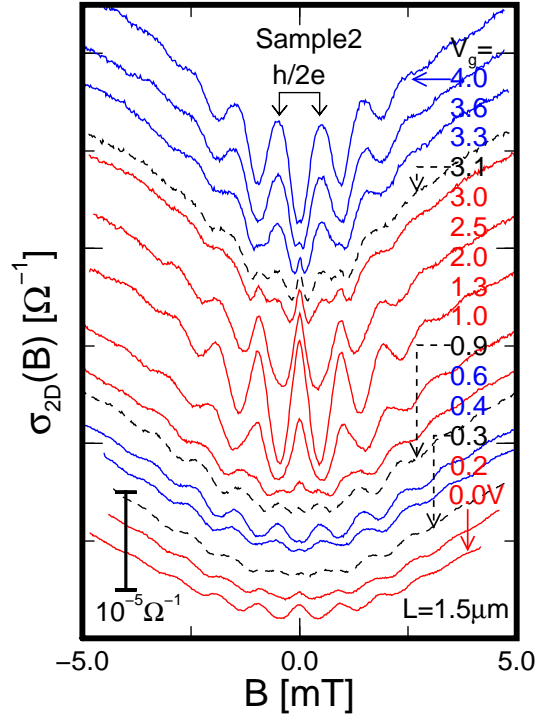}
\vspace{-2em}
\caption{\label{fig:sigma_raw_data} Gate voltage dependence of the 
electric sheet conductivities $\sigma_{\rm 2D}$ as a function of the 
magnetic field $B$ for a square loop (SL) array sample ($L=1.5\mu$m) 
fabricated using the sample2 epi-wafer in 
Ref.~\onlinecite{koga02wal}. The plotted curves are shifted along 
$y$ axis for the ease of comparison. The magnitudes of 
$\sigma_{\rm 2D}$ at $B=0$ range from 3.7$\times$10$^{-4}$ 
$\Omega^{-1}$ (for $V_{\rm g}=0.0$ V) to 10.3$\times$10$^{-4}$ 
$\Omega^{-1}$ (for $V_{\rm g}=4.0$ V). The range of $B$ ($\Delta B$) 
that corresponds to the magnetic flux half quanta piercing the SL 
($\Delta B\times L^2=h/2e$) is indicated by ``$h/2e$'' in the figure.}
\end{figure}

\begin{figure*}
\includegraphics[width=7cm,clip,keepaspectratio]{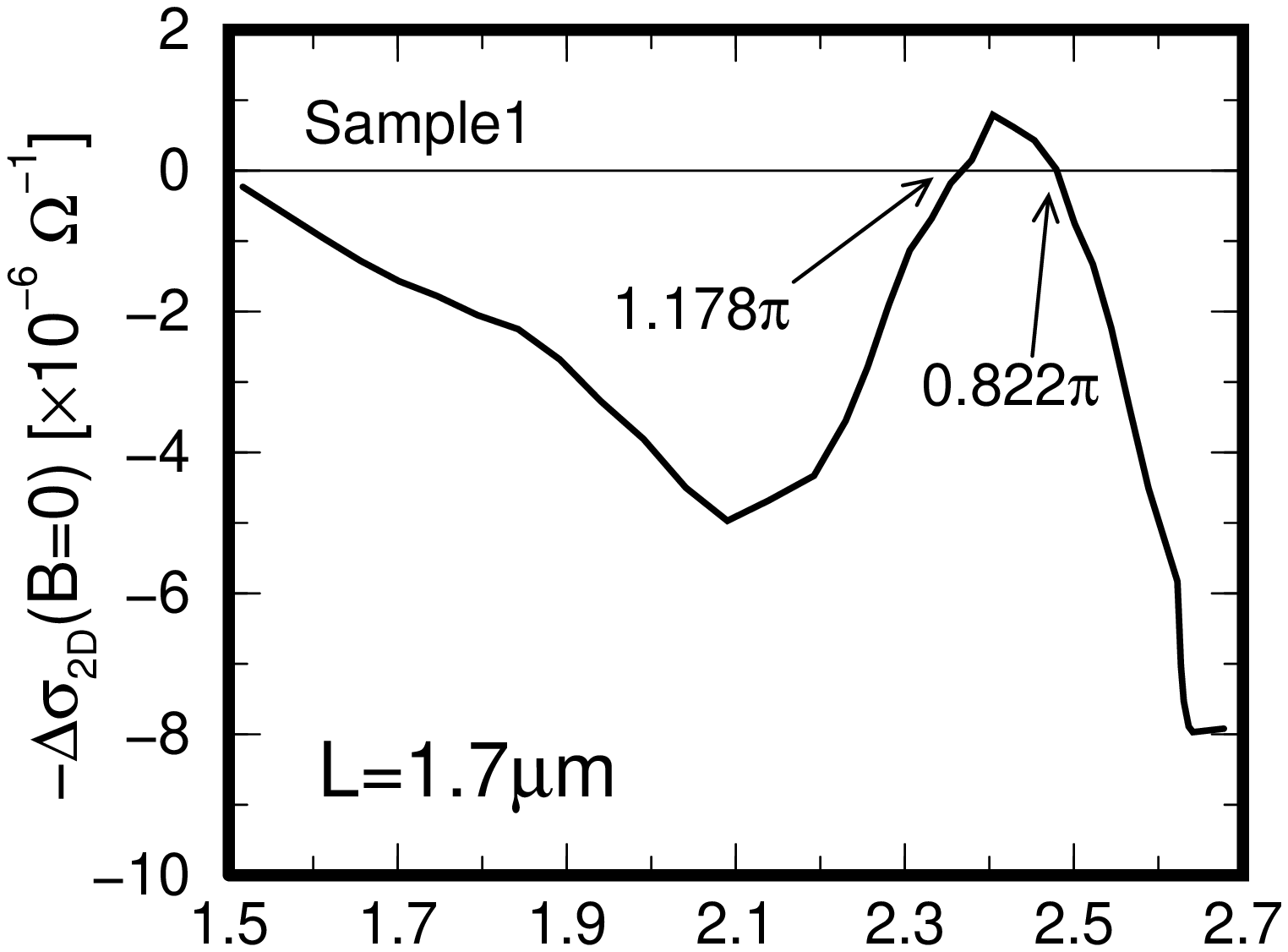}
\includegraphics[width=6.4cm,clip,keepaspectratio]{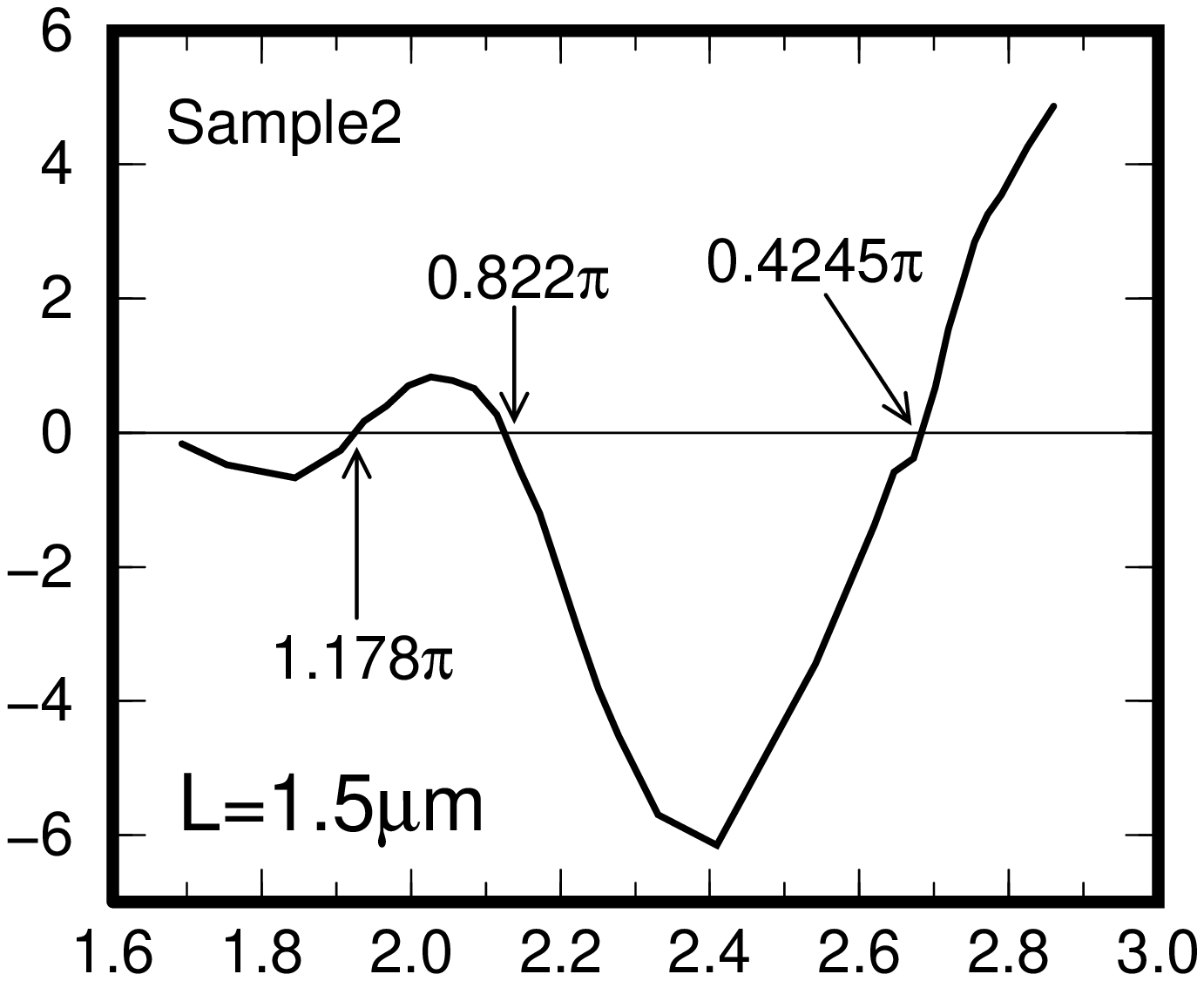}\\
\includegraphics[width=7cm,clip,keepaspectratio]{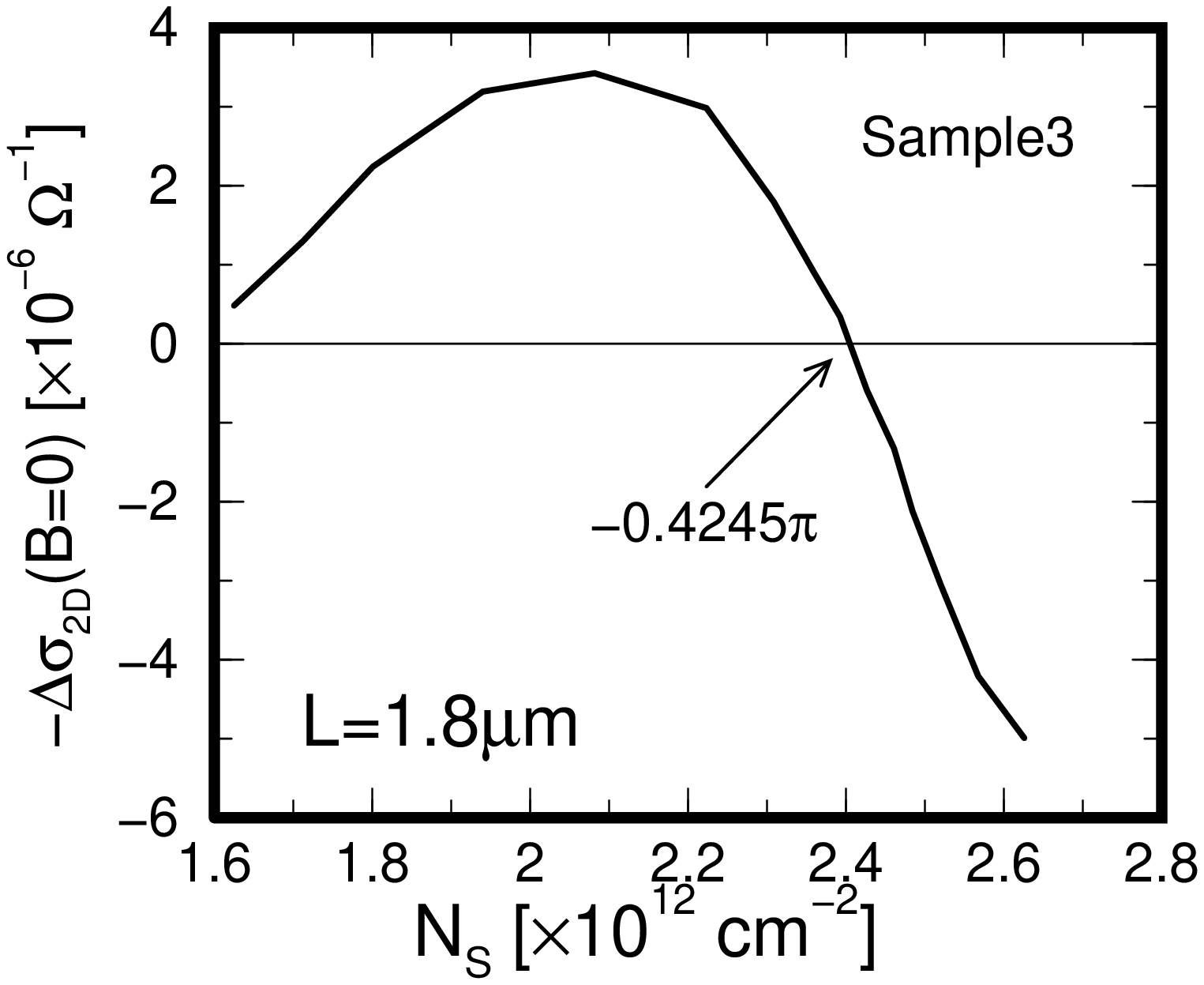}
\includegraphics[width=6.4cm,clip,keepaspectratio]{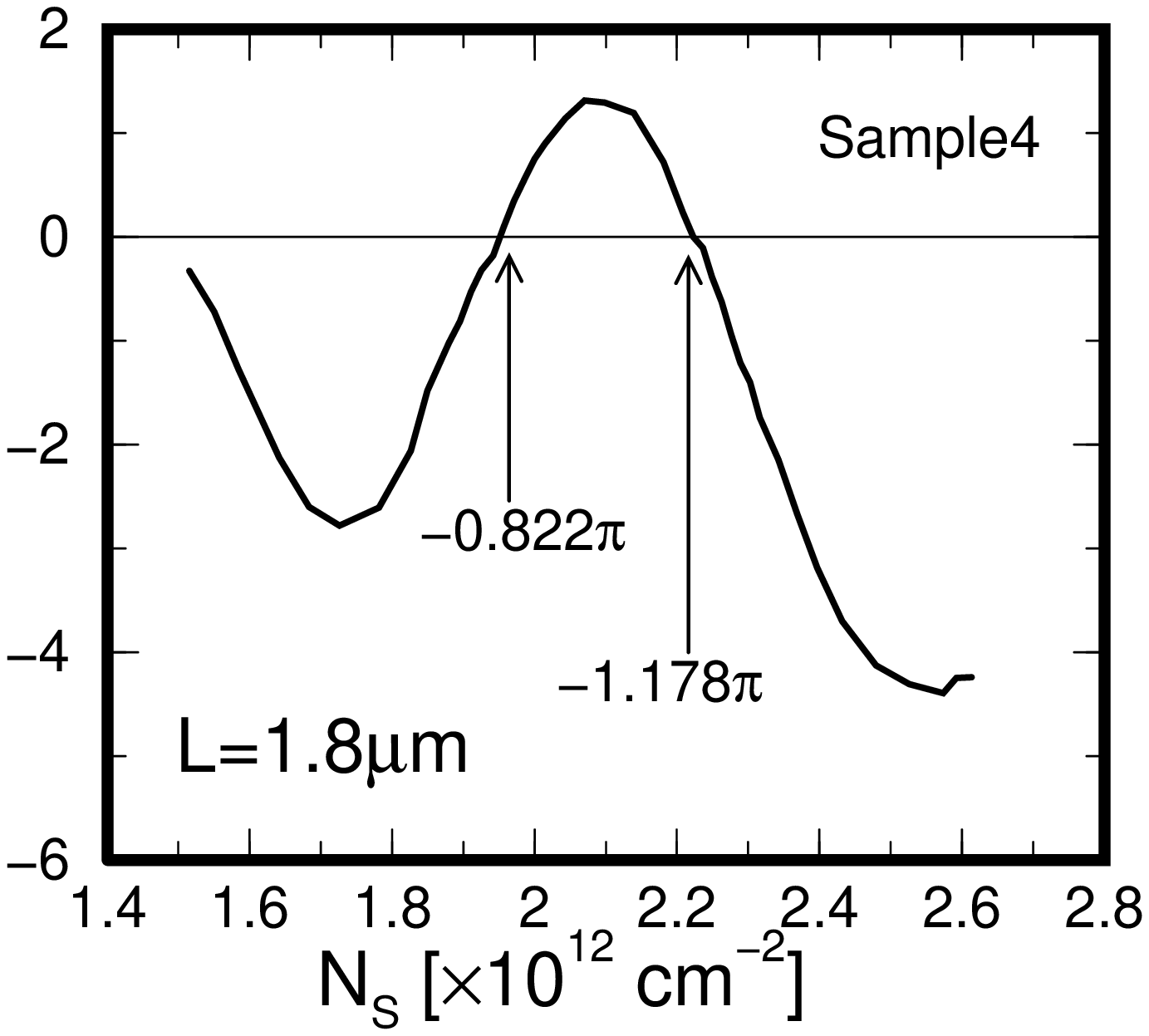}\\
\vspace{-2em}
\caption{\label{fig:AMP} Amplitudes of the experimental 
AAS oscillations at $B=0$ measured for various SL array samples ($L=1.5-1.8$ 
$\mu$m using the sample 1$-$4 epi-wafers introduced in 
Ref.~\onlinecite{koga02wal}) plotted as a function of the 
sheet carrier density $N_{\rm S}$. $\theta$ values at the node positions 
(denoted as $\theta^*$ in the text) are also given. We plot 
$-$$\Delta\sigma (B=0)$ instead of $\Delta\sigma (B=0)$ to match the signs 
of the values with those for $A(\theta)$ given in Eq.~(\ref{eq:A_theta}).
}
\end{figure*}

\begin{figure*}
\includegraphics[width=7cm,clip,keepaspectratio]{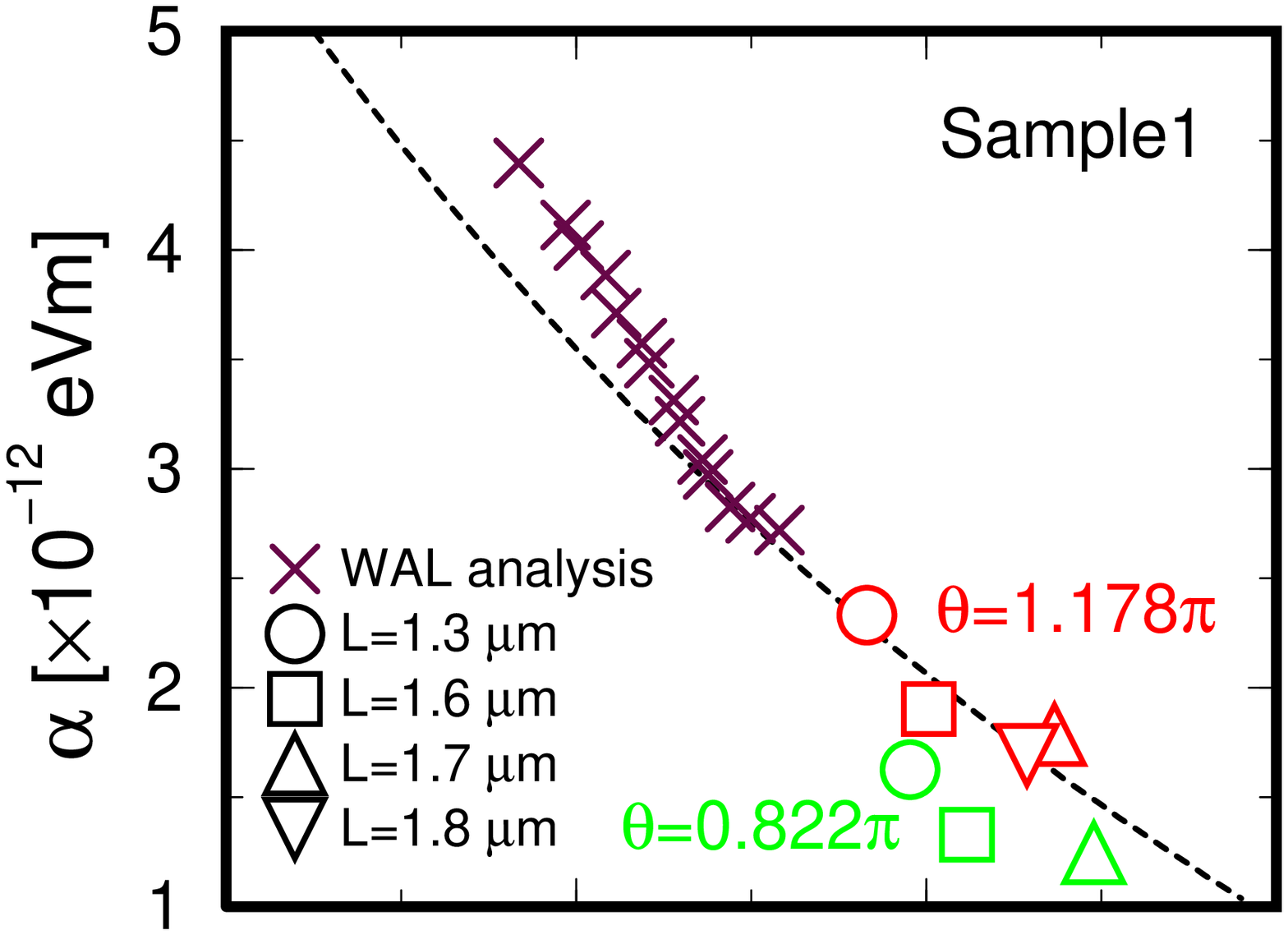}
\includegraphics[width=6.45cm,clip,keepaspectratio]{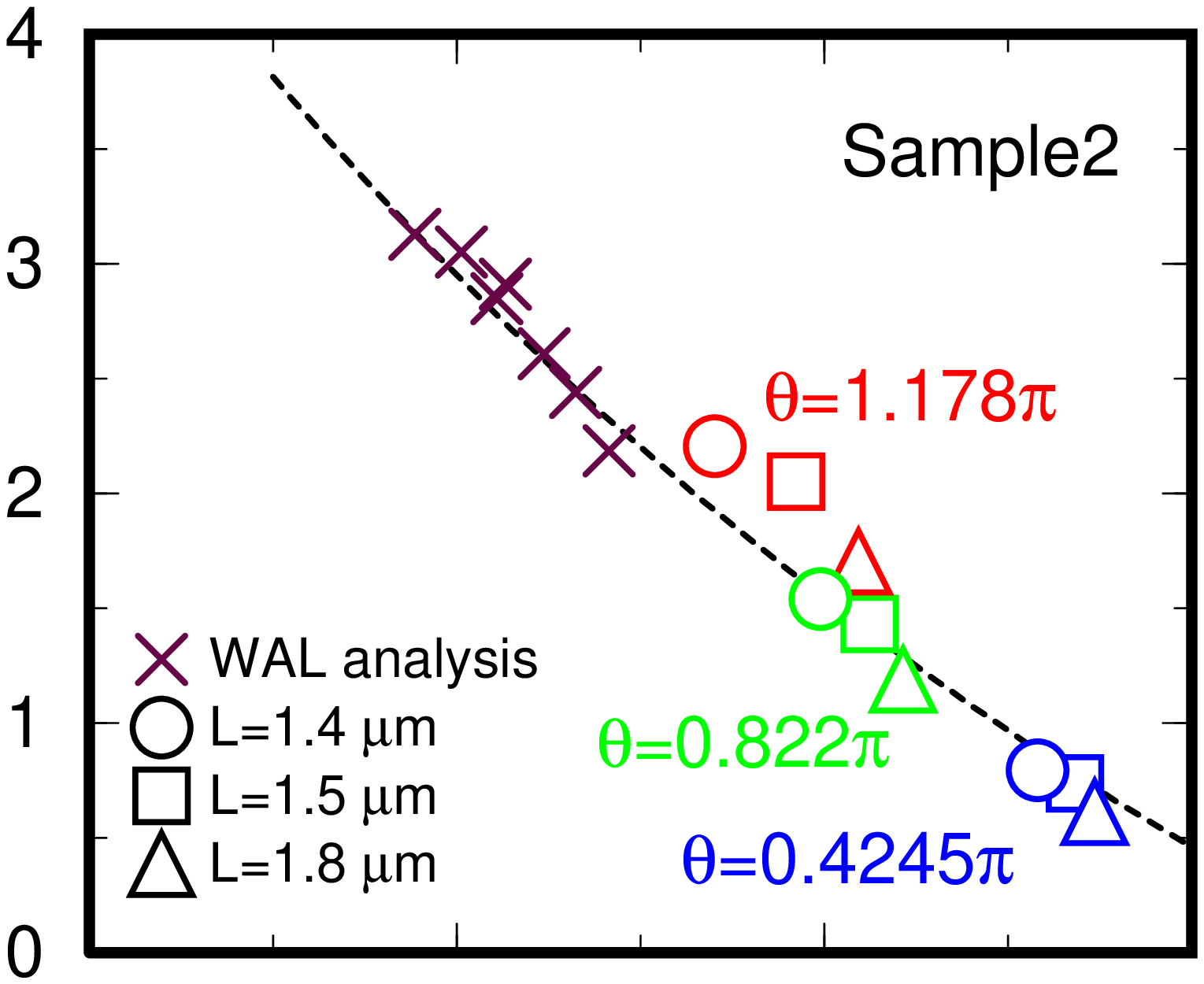}\\
\includegraphics[width=7cm,clip,keepaspectratio]{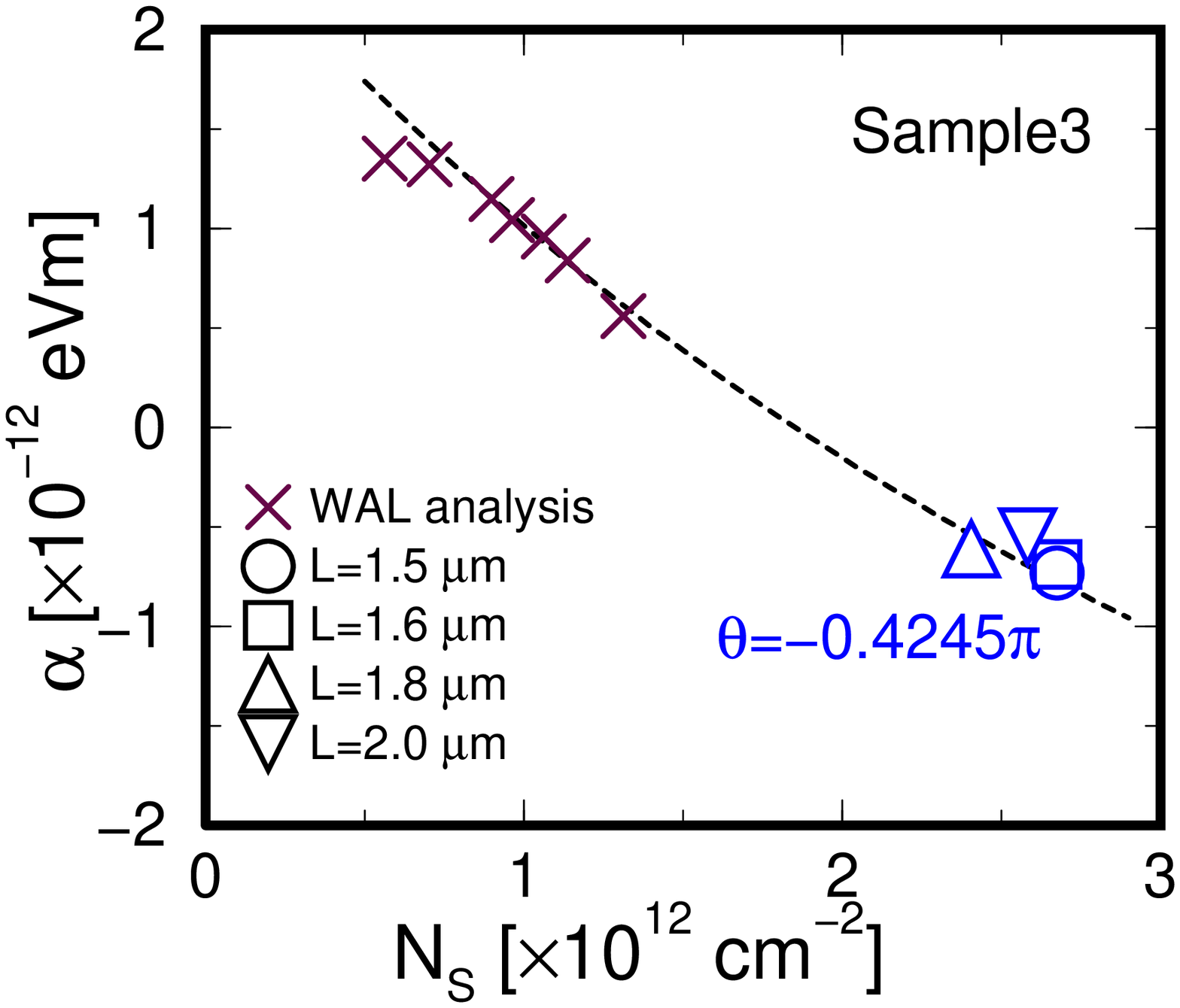}
\includegraphics[width=6.45cm,clip,keepaspectratio]{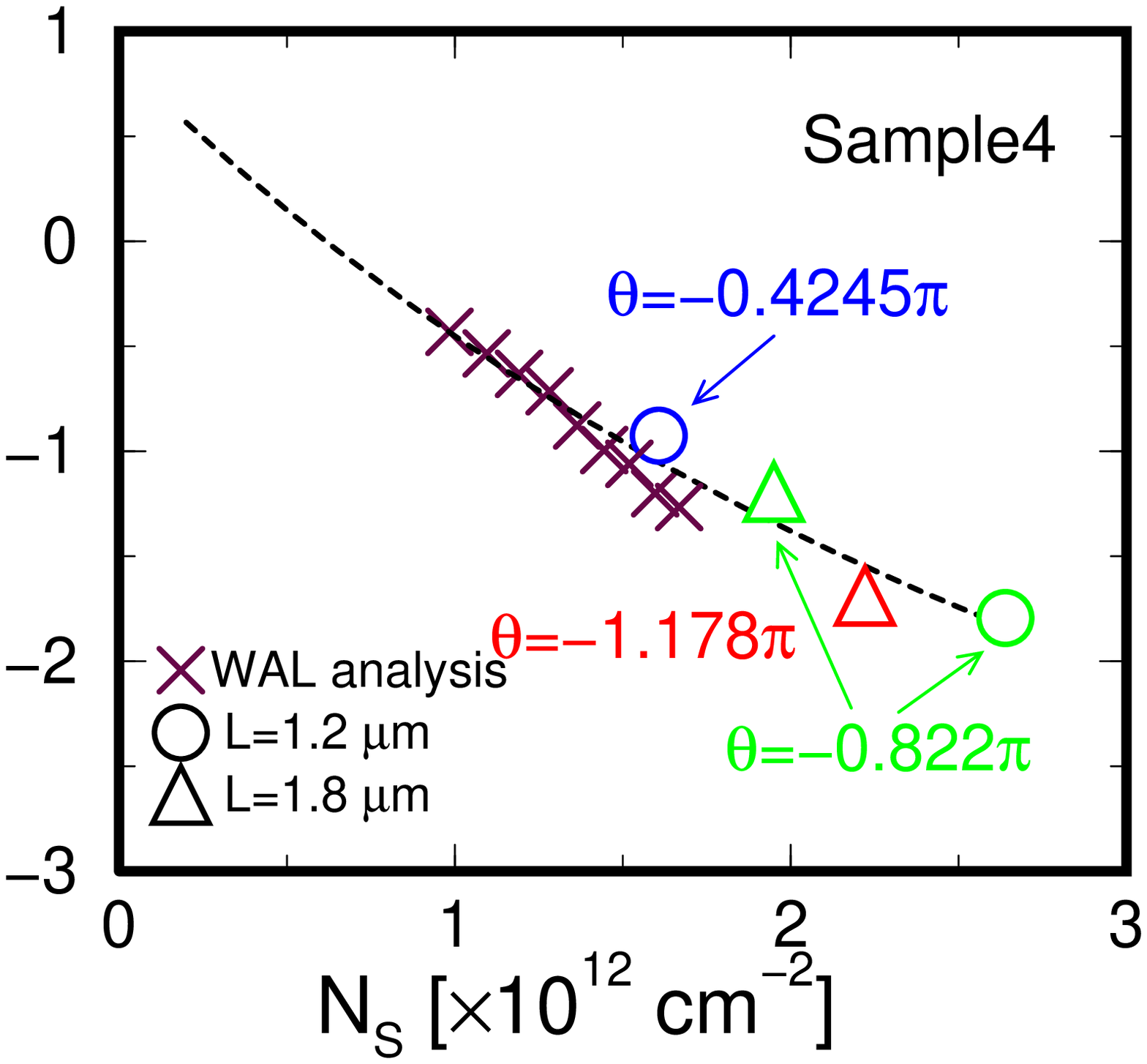}\\
\vspace{-2em}
\caption{\label{fig:alpha} The values of the Rashba spin-orbit parameter 
$\alpha$, for four different epi-wafers denoted as samples1-4 in 
Ref.~\onlinecite{koga02wal}, deduced from the three independent analyses: 
(1) the weak antilocalization analysis (crosses), (2) the analysis of the node 
positions in the $-\Delta\sigma_{\rm 2D}(B=0)$ vs. $N_{\rm S}$ relations for 
the square loop arrays using the relation $\alpha=\theta\hbar^2/2m^*L$
(various symbols) and (3) the ${\bf k}\cdot{\bf p}$
model calculations using appropriate boundary conditions (dashed curves). The 
background impurity densities ($N_i$) assumed for the ${\bf k}\cdot{\bf p}$ 
calculations are $N_i=1\times 10^{16}$, $4\times 10^{16}$, $1.4\times 10^{16}$ and 
$1\times 10^{16}$ cm$^{-3}$ for samples 1$-$4, respectively.}
\end{figure*}

Shown in Fig.~\ref{fig:sigma_raw_data} is the gate voltage 
($V_{\rm g}$) dependence of $\sigma_{\rm 2D}(B)$ for a SL array sample 
($L=1.5\mu$m) that is fabricated using the sample2 epi-wafer in 
Ref.~\onlinecite{koga02wal}. Here, we clearly see the AAS 
oscillations, whose 
period ($\Delta B$) is given by $h/2eL^2$.
We also note that as the value of $V_{\rm g}$ is increased from 
0.0 V, the peak feature in $\sigma_{\rm 2D}(B)$ at $B=0$ become dip 
across  $V_{\rm g}=0.3$ V [a dashed $\sigma_{\rm 2D}(B)$ curve]. 
Then, the dip feature becomes peak for $V_{\rm g} > 0.9$ V [also 
indicated by another dashed $\sigma_{\rm 2D}(B)$ curve]. Finally 
the peak feature again becomes dip for $V_{\rm g} > 3.1$ V. Thus the 
amplitudes of the AAS oscillations at $B=0$ oscillate as a function of 
$V_{\rm g}$ as predicted in Eq.~(\ref{eq:A_theta}).

Plotted in Fig.~\ref{fig:AMP} are the amplitudes of the experimental 
AAS oscillation at $B=0$ [denoted as $\Delta\sigma_{\rm 2D}(B=0)$] as a 
function of $N_{\rm S}$ for the SI devices fabricated using the sample 
1$-$4 epi-wafers ($L=1.7$ and 1.5 $\mu$m for samples 1 and 2, respectively, 
and $L=1.8$ $\mu$m for samples 3 and 4), where we employed the Fast Fourier 
Transform (FFT) and inverse FFT techniques to extract only the 
oscillatory part of $\sigma$ whose period corresponds to the magnetic flux 
half quanta $h/2e$. We indeed see that $-$$\Delta\sigma (B=0)$ oscillates 
with $N_{\rm S}$, where we observe several nodes. Using the $\alpha$ 
vs.~$N_{\rm S}$ relations that are obtained from the WAL analysis of an 
unpatterned QW sample and the ${\bf k}\cdot{\bf p}$ model calculation 
using appropriate boundary conditions \cite{koga02wal}, $\theta$ values 
for sample 2 at these node positions [denoted as $\theta^*$ below], for 
example, are identified as (from left to right) 1.178$\pi$, 0.822$\pi$ and 
0.4245$\pi$ (see Fig.~2 in Ref.~\onlinecite{koga04}). We thus demonstrated 
that the spin precession angle $\theta$ is gate-controllable by more than 
0.75$\pi$ for a length of 1.5$\mu$m. The $\theta^*$ values for the other SI 
devices using the other epi-wafers are also identified in Fig.~\ref{fig:AMP}.
We can, then, calculate the $\alpha$ values at these node 
positions using the relation $\alpha=\theta^*\hbar^2/2m^*L$. 

In Fig.~\ref{fig:alpha}, we plot the $\alpha$ values obtained in this 
way (denoted as $\alpha_{\rm SI}$) for various SL array samples made of the 
sample1-4 epi-wafers as a function of $N_{\rm S}$. Also plotted in 
Fig.~\ref{fig:alpha} are (1) the $\alpha$ values obtained from the WAL 
analysis of the unpatterned (bare) Hall bars (denoted as $\alpha_{\rm WAL}$)
and (2) those obtained from the ${\bf k}\cdot{\bf p}$ 
model calculations (denoted as $\alpha_{{\bf k}\cdot{\bf p}}$) using the 
appropriate boundary conditions and assuming the presence of the background 
impurities \cite{koga02wal}. We note that the unpatterned Hall bars for 
$\alpha_{\rm WAL}$ are prepared on the same wafer pieces as those used 
for the SL array samples. We also note that in Ref.~\onlinecite{koga02wal} 
we obtained $\alpha_{{\bf k}\cdot{\bf p}}$ values without assuming the 
background impurities and found quantitatively good agreement with 
$\alpha_{\rm WAL}$ values. In the present work, we included the effect of 
the background impurities (mostly they are present in the 
In$_{0.52}$Al$_{0.48}$As buffer layer) in the model calculation of 
$\alpha_{{\bf k}\cdot{\bf p}}$ to better fit the experimental 
$\alpha_{\rm WAL}$ and $\alpha_{\rm SI}$ values. It turned out that the 
values of the background impurity densities obtained from these fittings 
are reasonably small (typically 1$\times$10$^{16}$ cm$^{-3}$). The details 
of this analysis are discussed elsewhere \cite{sekine05}.

In summary, we have demonstrated experimentally the electron spin interference 
phenomena based on the Rashba effect, which are predicted previously 
\cite{koga04}. 
For this demonstration, we prepared nanolithographically defined square loop 
array structures in 
In$_{0.52}$Al$_{0.48}$As/In$_{0.53}$Ga$_{0.47}$As/In$_{0.52}$Al$_{0.48}$As 
quantum wells using the electron beam lithography and ECR dry etching 
techniques 
and measured the low-field magnetoresistances of these samples
(${\bf B}\perp$ sample surface) at low temperatures (0.3 K). We observed the 
Al'tshuler-Aronov-Spivak (AAS) oscillations, whose magnitudes at $B=0$ 
oscillated as a function of the gate voltage as the result of the 
spin interference. We also deduced the $\alpha$ values (Rashba spin-orbit 
coupling constant) from the analysis of the spin interferometry 
experiments. We obtained quantitative agreements among (1) the $\alpha$ 
values obtained from the spin interferometry experiments, 
(2) those obtained from the weak antilocalization analysis, and (3) those 
obtained from the ${\bf k}\cdot{\bf p}$ model calculations. 


\end{document}